# CAN THE PHOTOSYNTHESIS FIRST STEP QUANTUM MECHANISM BE EXPLAINED?

*Marco Sacilotti*[a,c], *Euclides Almeida*[a], *Cláudia C. B. O. Mota*[a], *Frederico Dias Nunes*[b], *Anderson S. L. Gomes*[a]

[a] Department of Physics. Universidade Federal de Pernambuco, Av. Prof Luiz Freire, s/n, Cidade Universitária 50670-901, Recife, PE, Brazil. Phone: 55-81-21267636 Fax: 55-81-32710359.

[b] Department of Electronics and Systems, Universidade Federal de Pernambuco, Rua Acadêmico Hélio Ramos, s/n, Cidade Universitária 50740-530, Recife, PE, Brazil. Phone: 55-81-2718210 Fax: 55-81-2718995.

[c] CMN Group UFR Sc. Techn. FR 2604 – Université de Bourgogne, 9 avenue A. Savary, BP 47870, 21078 Dijon Cedex, France. Phone: 33-3-80395908 Fax: 33-3-80396013.

**Corresponding author:**

Marco Sacilotti, professeur émérite

ICB – UMR 5209 / FR2604 CNRS Université de Bourgogne

9, Av. Alain Savary

BP 470870 - 21078 Dijon France

Phone 33 3 80395809    Fax 33 3 80396013

E-mail: msacilot@u-bourgogne.fr & msacilot@gmail.com

**E-mails:**

Marco Antonio Sacilotti: msacilot@u-bourgogne.fr

Euclides César Almeida: cesarlins@df.ufpe.br

Claudia Cristina Brainer de Oliveira Mota: claudiabmota@gmail.com

Frederico Dias Nunes: nunes.fd@gmail.com

Anderson Stevens Leônidas Gomes: anderson@df.ufpe.br

**Corresponding author:**

Marco Sacilotti, professeur émérite

ICB – UMR 5209 / FR2604 CNRS Université de Bourgogne

9, Av. Alain Savary

BP 470870 - 21078 Dijon France

Phone 33 3 80395809    Fax 33 3 80396013

E-mail: msacilot@u-bourgogne.fr & msacilot@gmail.com

**E-mails:**

Marco Antonio Sacilotti: msacilot@u-bourgogne.fr

Euclides César Almeida: cesarlins@df.ufpe.br

Claudia Cristina Brainer de Oliveira Mota: claudiabmota@gmail.com

Frederico Dias Nunes: nunes.fd@gmail.com

Anderson Stevens Leônidas Gomes: anderson@df.ufpe.br

**Abstract**

Photosynthesis first step mechanism concerns the sunlight absorption and both negative and positive charges separation. Recent and important photosynthesis literature claims that this mechanism is quantum mechanics controlled, however without presenting qualitative or quantitative scientifically based mechanism. The present accepted and old-fashioned photosynthesis mechanism model suffers from few drawbacks and an important issue is the absence of driving force for negative and positive charges separation. This article presents a new qualitative model for this first step mechanism in natural catalytic systems such as photosynthesis in green leaves. The model uses a concept of semiconductor band gap engineering, such as the staggered energy band gap line-up in semiconductors. To explain the primary mechanism in natural photosynthesis the proposal is the following: incident light is absorbed inside the leaves causing charges separation. The only energetic configuration that allows charges separation under illumination is the staggered one between two materials or molecules. We explain why (e-, h+) interacting charges can be separated by using an energy staggered configuration, under illumination. Following this model, the green light of plants can be seen as related to the spent energy for charges separation. Green colour being mostly an emission and not a reflection as currently presented. The arguments mentioned below show why we cannot explain the photosynthesis first step mechanism, based on the presently accepted model, and therefore the alternative model is presented.

**Keywords:** photosynthesis, catalysis, type II interfaces, quantum photosynthesis, tunnelling, staggered energy band gap, energy, photosynthetic, solar energy, plants, leaves.

**Introduction**

Green, yellow, brown, red, or…? What was the colour of the Earth´s canopy at its very beginning or at its highest exuberance? What was the colour of the forests that created the fossil petroleum that we are consuming today? We can, maybe, say that it was red shifted from the present colour, if we accept the idea that the plant's green colour is the spent energy by the first step mechanism of photosynthesis for the charges separation. If ($h\nu_{green}$ - $h\nu_{red}$)/$h\nu_{green}$ represents around 29% of energy economy, the canopy red emission should represent a Nature's better performance compared to the green emission we have today. This paper intends to propose and to discuss a different model for the first step mechanism of photosynthesis, based on Nature's existing band energy engineering. It intends also to show how and why nature's machinery can use sunlight more efficiently by its own physical and natural quantum mechanical choices. The present proposal will be presented in such a way to be understood by biologists, chemists, physicists and amateurs interested on the photosynthesis mechanism. Presented without equations and the difficult statements of quantum mechanical physics, it is required of the reader knowledge of quantum tunnelling of charges (wavefunction overlapping) and electrical charge sitting in a quantum well (barriers). The proposed band gap energy engineering can be seen as a fingerprint of different species (materials or molecules) band energy by its relative position.

Photosynthesis is the biological process in which green plants use solar energy to reduce $CO_2$ and oxidize $H_2O$. The primary events of this quantum conversion process (by absorbing a photon – a quantum of light energy) involve the absorption of photons (light) by the molecules (e.g., chlorophyll, carotene, etc) with subsequent charge separation in the chloroplast. It is not known exactly how the pigments, proteins, and their environment are organized to facilitate the quantum conversion process (a photon energy transformed into charges separation mechanism). In summary, the first step process of photosynthesis is not known exactly. It can even be debated if leave's green colour is mostly an emission or mostly a reflection.

Theoretical calculations for the photosynthesis first step process of energy bands representing most of the molecules species band energy physical parameter's (in green leaves) suffers from a unique and basic statement: the ground state energy (or the bottom and common energy level for non excited charges). It is a supposed common energy level for all steps considered for energy transfer in photosynthesis[1-5]. Researchers admit that the ground state energy is "uncalibrated" (see fig. 2 of ref. 6). If so, this is not a physically acceptable situation for interacting physical systems. If ones accept that do exist a common ground state energy for different molecules inside a leaf, another physical problem arises: where is the origin of the driving force for interacting (e-, h+) charges separation? Without an electric field, the driving force is absent and there is no longer charges' separation (see fig. 12 of ref. 7). Another unsolved problem within the present first step of photosynthesis mechanism is the usual removal of the chemical constituents of leaves from their natural environment to measure optical properties, such as absorption and fluorescence[2]. The above mentioned arguments show why we cannot explain the photosynthesis first step by using the presently accepted model.

Natural species as a leaf has chemical environment and physical properties (size, distances and composition) that can change the optical properties of their individual constituents. So, most of the physical/chemical processes utilized to study natural plants (as chemical separation by ultracentrifugation or isolation procedure) do not allow researchers to get a real picture of the natural, *in vivo*, photosynthesis mechanism[2, 5]. Considering the bottom common energy line (or the ground state energy) and considering the leave's constituents (chemical and physical) as separated components to develop a real photosynthesis mechanism is a naive physical expectation. So if the relative position of the band gap energies of the leaf's constituents is unknown, it must be calibrated. In doing so, both (e-, h+) charges undergo energetic steps when going from one material to another. These energetic steps are very important for the optical properties, for the charges separation and the subsequent processes in plants. Moreover, the flat energy band representation when excited materials have electrical charges jumps is a physically incorrect situation and model [7, 21, 39].

**The staggered band gap configuration**

Based on semiconductor or inorganic material band-gap engineering, a common energy baseline for a pair of materials is a very rarely encountered physical situation[7-12]. Ignoring lattice parameter matching, it is clear that any two semiconductors with the same energy band gap could be non-staggered only by accident[10].

Semiconductor band gap engineering has been one of the most important tools for electro-optical devices and it is the key for the huge amount of development of optoelectronics/photonics since the 70's. Within these devices pairs of organized materials and their band gaps adjusted such that one narrow band gap fits energetically within the wide band gap of an adjacent material. These are called type I interfaces[7-12]. But most energy band gap engineers have ignored a less utilized energy band gap alignment, where both band gaps are staggered, one related to the other. These are defined as type II energetic interfaces. In the 70's and 80's, two Nobel laureates (L. Esaki, 1973, H. Kroemer, 2000) worked on many kinds of semiconductor energy band gap alignments[7-13]. One such alignment is the type-II or staggered interface that we describe within the following lines. The appearance of band offsets creates energetic steps or electric potential steps. This gives rises to barriers for e- and

h+ at the interface. In this case and near the interface, most of the photoelectrons (e-) go to the lower energy step conduction band (CB) and most of the holes (h+) go to the lower energy step valence band (VB). In this way, the energy band-bending near the interface for e- creates an exponential-like quantum well in the CB of one material. The energy band-bending near the interface for h+ creates an exponential-like quantum well in the VB of the other material. Between these wells exists the physical interface among the two materials and potential barriers. The energy band bending on each side of the interface gives rise to the necessary electric field for the charges separation (note that the electric field **E** = **-grad** V, V = electrostatic potential and Energy = V. charge). The charges of opposite signals are at different materials (e-, h+), separated by the interface. However they can recombine when tunnelling effect is considered, according to their wavefunction overlap (or tunnelling)[10,12,13]. As (e-, h+) are located in different materials, the recombination occurring at the interface takes place free of any quantum mechanical **k** selection rule (**k** = momentum), following the lack of periodicity. This fact improves the probability of recombination making it a more efficient recombination/emission mechanism. Calculation and representation on the (e-, h+) wave function barriers penetration are presented on support information (SI-1). Recently many workers mention that in their material systems, charges are located in different materials when they recombine or they are at the material's surface, when (e-, h+) recombine[14,15]. In the same way, oxides or surface states can be seen as charge-trapping centres that keep one of the carriers (e-, h+) outside the crystal lattice[16-20].

We propose in this paper that these (e-, h+) recombination mechanisms of many unsolved natural and/or artificial systems can be interpreted as type II interface recombination involving (e-, h+) charges located at different materials. This paper will present formerly published type II interface band gap engineering for one III-V semiconductor AlInAs/InP pair[10,12,21]. It is proposed a (e-, h+) separation/recombination/emission mechanism applied to natural photosynthesis first step, based on staggered band gap energy alignment.

**Staggered band gap engineering for the AlInAs/InP system**

Based on published data, (e-, h+) pairs (or electron/hole production by light absorption) can be furnished by type II staggered semiconductor band gap energy alignment[7, 12, 21]. This type II semiconductor interface is proposed below as an explanation for the primary mechanism in photosynthesis, without going into detail for the subsequent steps in the total process.

In figure 1, materials A (AlInAs) and B (InP) can be grown easily by the chemical vapour deposition process[12,13,21]. During the growth process, a sharp interface is formed between both A/B materials. This interface is staggered in energy band gap according to theoretical predictions and experimental results[7,10,12,13,21]. The optical emission properties at room temperature or at 77 K of this A/B materials shows a broad photoluminescence (PL) energy peak near 1.2 eV, as can be observed in figure 1.

*Figure 1*

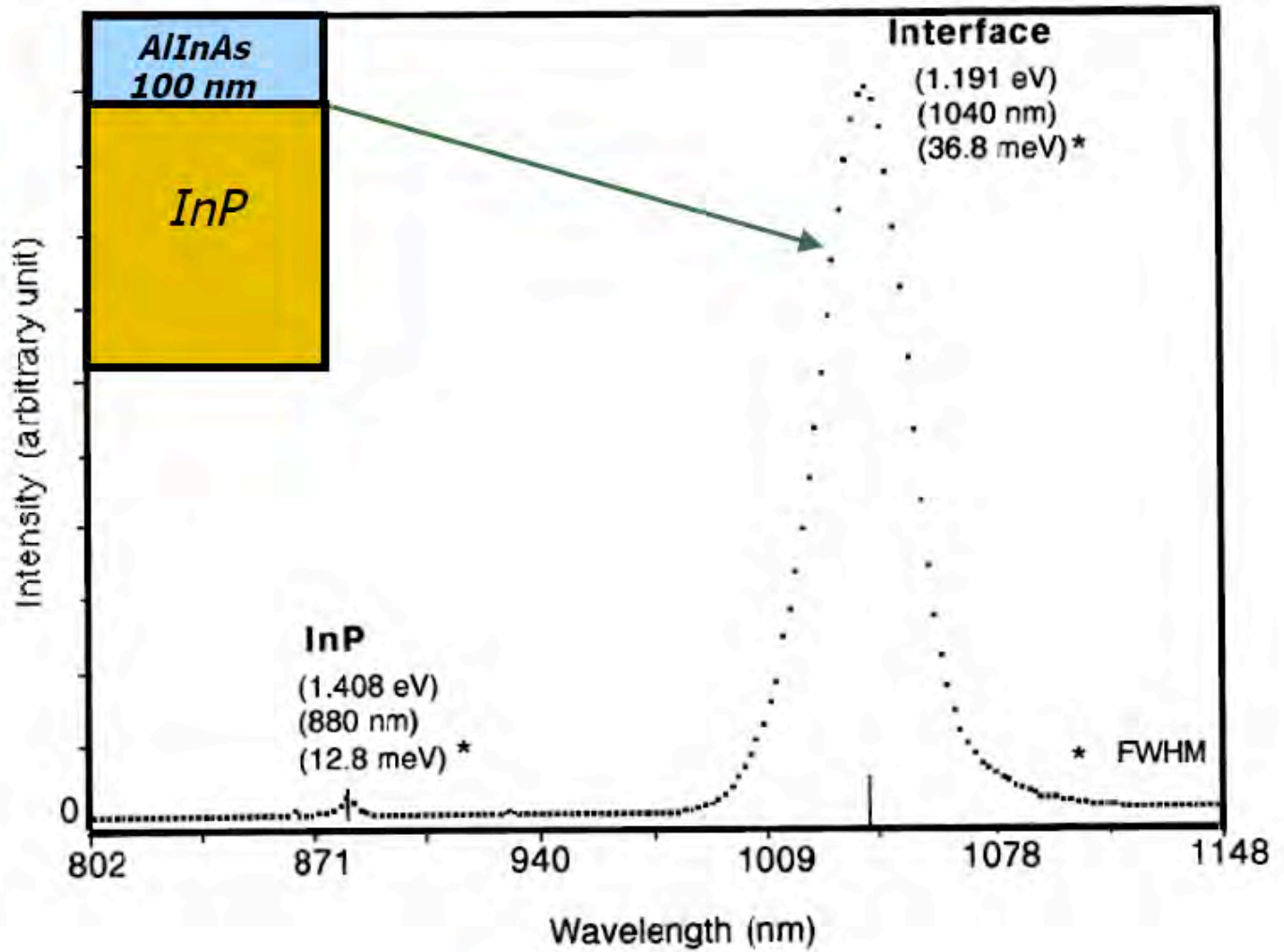

**Figure 1.** Type II semiconductor structure for material A (AlInAs) and B (InP), and its 77K photoluminescence (PL) spectrum for the AlInAs (100 nm tick) grown on InP by MOCVD [12, 21]. This spectrum shows a high intensity peak emission at 1.191 eV energy, originated from the AlInAs/InP interface. No PL peak is present for the AlInAs (829 nm) layer. A very weak intensity emission is present for the InP (879 nm) buffer layer and substrate. The broad FWHM = 36,8 meV value is a characteristic of type II interface emission.

As presented in references 10, 12, 13 and 21, this energy emission peak comes from the (e-, h+) recombination at the AlInAs/InP interface. As the argon laser light excites both A and B materials, (e-, h+) carriers are created on both sides. These carriers have diffusion lengths of the order of 1 μm (much higher than the AlInAs layer thickness of figure 1). Both (e-, h+) in the conduction (CB) and valence (VB) band have energetic steps (up and down) at the A/B interface. Due to the (e-, h+) decaying to lower energies at the interface, an interface band bending (as presented in figures 2 and 3) is created during the argon laser illumination. This band bending or band energy curvature depends on the number of (e-, h+) decaying[7-10, 12, 13, 21]. The band bending also creates a potential barrier against the decay of more carriers on the band offsets. These bands bending barriers are responsible in part for the electron and hole separation from the interface. Thus, part of the excitation light energy appears as emission due to recombination at the interface, and part of the light absorbed energy is used to separate electron and holes charges away from the interface. This mechanism is represented in figures 2 and 3 for the interface emission and (e-, h+) separation by band bending at the interface of the A/B materials.

Figure 2

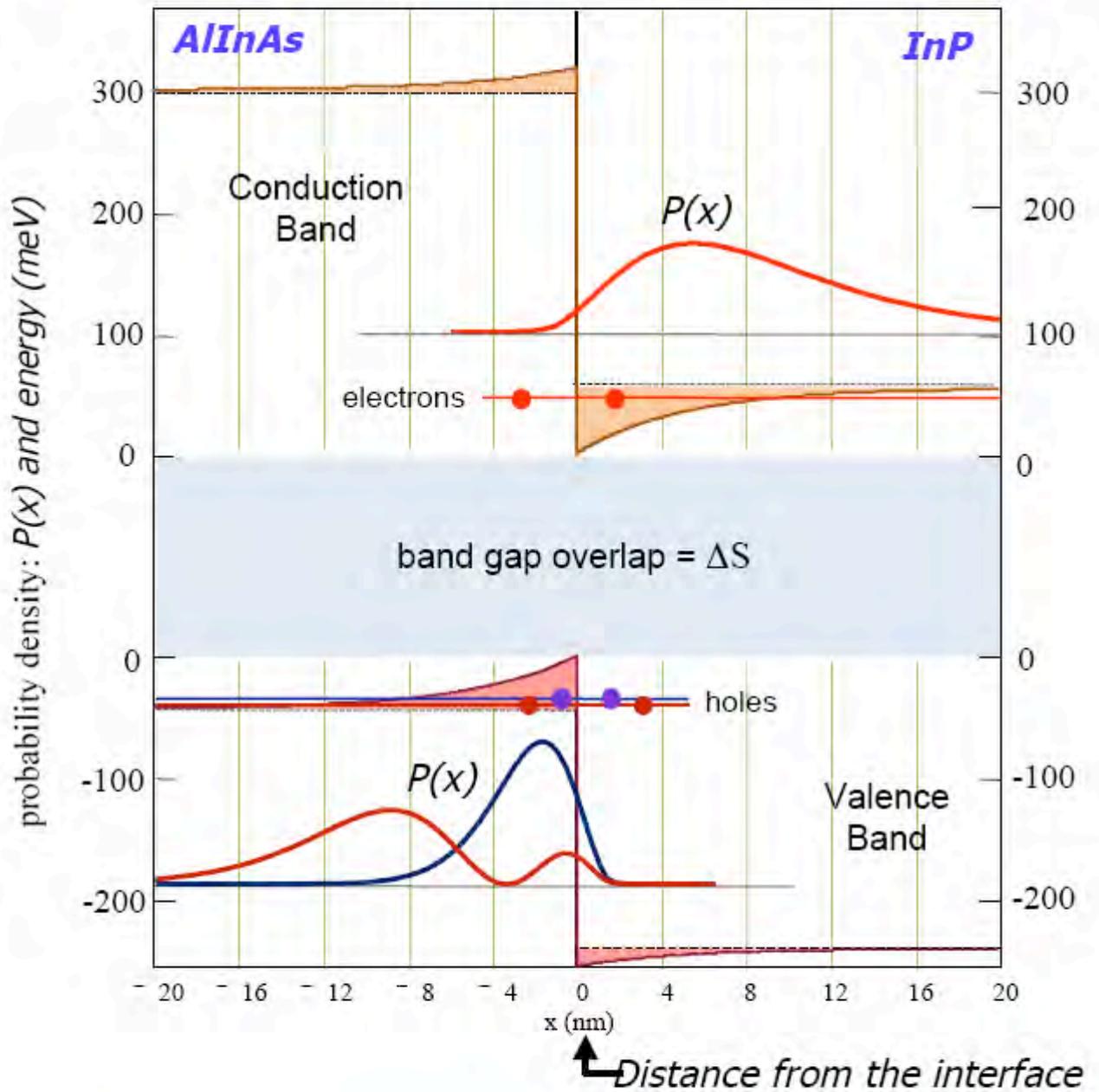

**Figure 2.** Picture representing the band energy bending at the interface, which is responsible by the (e-, h+) triangular quantum wells at both structure's sides. Included in this figure, near the AlInAs/InP interface, there are both representation of the (e-, h+) quantum mechanics wavefunction density of probability P(x), for the band offsets barrier's penetration: $\Delta E_c$ = 349 meV for electrons and $\Delta E_v$ = 272 meV for h+ barriers steps. P(x) represents the density of probability to found an electron or a hole at the position x within the structure. Note the e- and h+ P(x) overlap at the interface, position at x = 0. $\Delta S$ = material A/B band energy relative position overlap. There are two densities of probabilities for h+ and one for e-, regarding the conditions and parameter utilised for the calculations (see support information for more details for these calculations, SI-1).

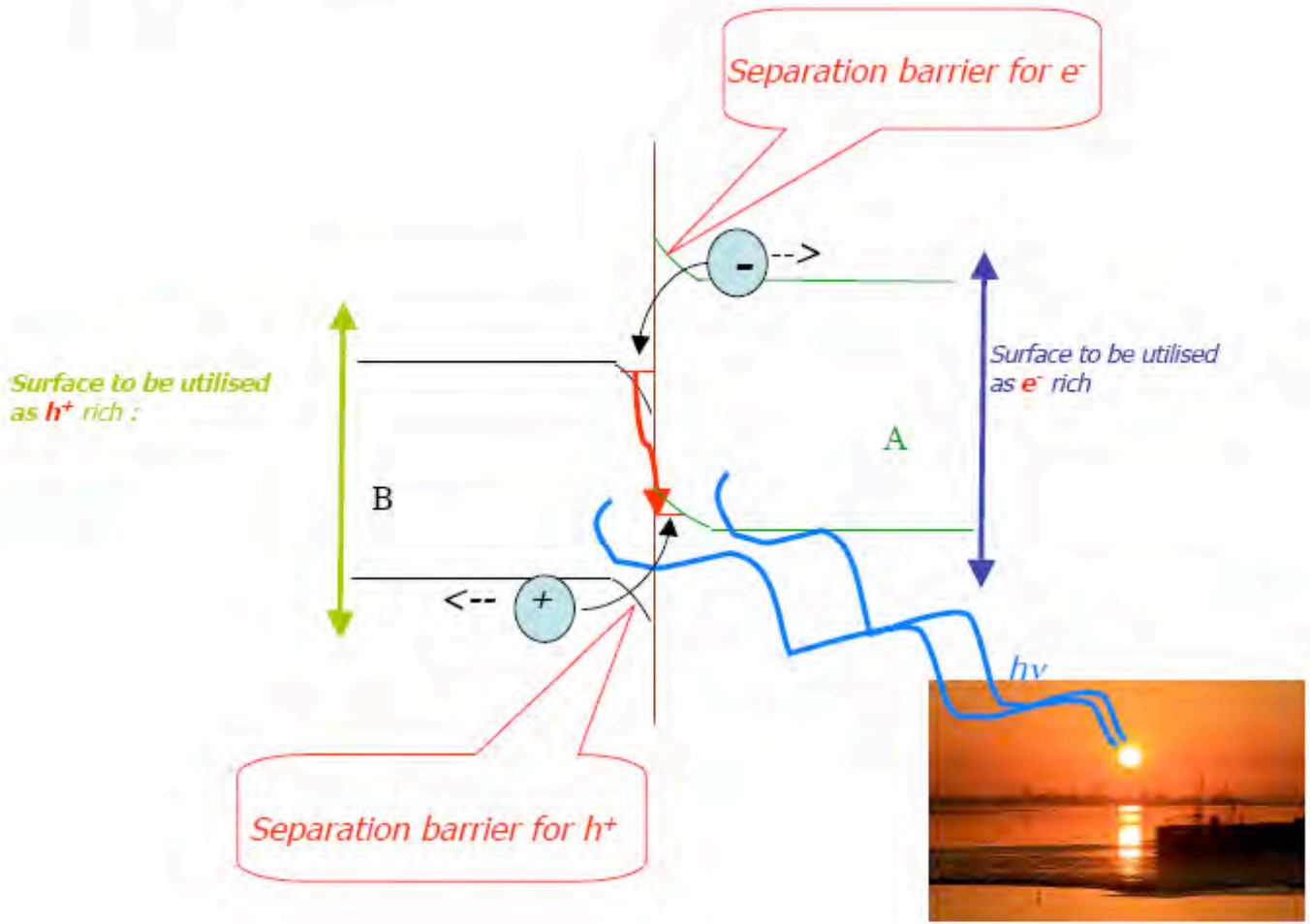

**Figure 3.** Band gap energy staggered line-up representation for the sunlight excited A/B material's structure. The dynamic excitation of this A/B structure creates exponential-like quantum wells at both sides: for e- (on A) and for h+ (on B). The decaying of e- and h+ from stepped barriers forming exponential-like quantum well allows energy band bending at each energy step. In a dynamic condition (by light excitation), these interface band bending are responsible by charge separation, going to the opposite side of the A/B interface. These separated charges can be responsible by catalytic effect like on $H_2O$ splitting (h+) and $CO_2$ reduction (e-). Part of the excitation energy is utilised for interface emission (like the 1.191 eV peak on figure 1), and the other part is conducted by the band bending to the A/B surface (opposite to the interface) to participate on both the oxidation and reduction catalytic processes. A = AlInAs, B = InP, CB = conduction band, VB = valence band.

There are two concurrent effects at the same time: decaying (e-, h+), that creates the band bending and, at the same time, separates others (e-, h+) charges from the interface. As some of the (e-, h+) pairs recombine at the interface and give rise to the 1.2 eV PL peak, it is related to the spent energy for the band bending (barriers) creation. Moreover, this energy (1.2 eV) is responsible for the charge separation mechanism (band bending creates barriers and the necessary electric field for). In this way, the interface recombination and emission gives rise to the charge separation in a dynamic process. This mechanism will be related to the emission of light by plants (the 1.2 eV peak is postulated as analogous to green light emission from leaves), and the band bending is responsible for the charge separation in the photosynthesis first step process. The dynamic process at the interface creates an exponential-like quantum well for h+ at the B side (InP) as well for e- at the A side (AlInAs). Physically this mechanism creates a dynamic electric field whose direction is from B to A. In other words, this electric field is responsible for the separation of both (negative and positive) charges. Based on band gap engineering, this is the only energetic configuration (type II interface) able to separate both (e-, h+) charges under illumination[7-10,12,13,21]. Moreover, this interface opposite quantum wells explains most of the recent literature conclusions claiming photosynthesis as a quantum mechanical mechanism[22]. SI-5 discusses about the recent literature conclusions on photosynthesis, including the present proposed mechanism.

Note that the energy band gap of both materials (A and B, of figure 1) are able to absorb most of the argon laser light (if hν is greater than or equal to the A and/or B band gap energies). The equation below represents approximately the interface light emission energy (hν), as stated in SI-2:

$$h\nu = \Delta S + \Delta Q_e + \Delta Q_h - \Delta E_x$$

Where: material B band gap energy > hν < material A band gap energy, $\Delta S$ = material A/B band energy relative position overlap, $\Delta Q_e$ = electron energy quantization within the exponential-like quantum well, $\Delta Q_h$ = hole energy quantization within the exponential-like quantum well, and $\Delta E_x$ = electron-hole excitonic interaction energy.

The above discussions and equations are for semiconductor pairs, as described in references 7-10, 12, 13, 21. Under light illumination with energy above the A and B band gaps, the dynamic band bending appears. If the illumination is turned off, the flat band condition comes back and no interface emission is present anymore. The band gap engineering, the charges wavefunction penetration and probabilities for figures 2 and 3 are presented in SI-1. Note that wavefunction penetration for electrons on AlInAs material is about 2 nm and the wavefunction penetration on InP for holes is about 1 nm (figure 2). These values are above or equal the dimensions and distances within natural leaves constituents. We performed some optical experiments on intact leaves to show that under excitation, they can present red shift emission. These results are presented on SI-3 and 4. When leaves are excited by violet light, we observe green light emission. Our results also show that the wavelength of light emitted by the leaves is dependent on the wavelength of excitation. On SI-4 we present one possibility among many ones that can exist within a leaf as a staggered configuration. This one possibility and the complexity of the task to find others possible staggered energy band gap give us an idea about the difficulties and why scientists seems to not have being able to appropriately explain the first step in the photosynthesis mechanism. It represents a real Mother's Nature puzzle.

**Discussion**

We proposed that, to the nature's green leaves colour, with a chemical/physical structure (chromophores), it is associated to type II semiconductor staggered energy interface mechanism. The leave's physical structure is composed of about 1 nm dimension for both physical structure and chemical elements (chlorophil1s, carotenoids etc). By the charges (e-, h+) wavefunction overlap and tunnelling, a mechanism of recombination/emission is present at the interface of a couple of materials, where both carriers are seated in different materials (no **k** quantum mechanical rules is required for this recombination). The decaying of (e-, h+) carriers to the interface, creating exponential-like quantum wells (deep and thin) create interface energy band bending at the same time (dynamic process). At both materials CB and VB there are the energy band bending of the energy valence band level and the band bending of the conduction band energy level. So part of the absorbed sunlight energy is transformed into emission (at the interface, corresponding to the ~1.2 eV for the A/B structure of figure 1). This is assumed to be the green colour of the leaves (a fast femto-second process). Part of the carriers that do not recombine at the type II interface are put far away from this interface by the action of the created electric field, due to this interface energy band bending. By the driving force, these electric charges go to the opposite side of the type II interface (slower process). These charges are responsible by the next step of the photosynthesis process to produce the plant's biomass and the water splitting.

Note that in figure 1, the type II interface related AlInAs/InP structure dimensions are 100 nm/1 μm. In plants, the dimensions of molecules are far below 10 nm. Band bending should be much more pronounced in this case. So the exponential-like interface quantum well deepness should be much more pronounced. The (e-, h+) wavefunction overlap should be much more efficient and recombination/emission occurs much more easily, without quantum mechanics selection rules, because both (e-, h+) are seated in different materials. As the deepness increases, it represents a broad band recombination/emission, as is the case for the AlInAs/InP interface emission and also for plant's leaves green emission (figure 1, SI-3 and 4 and references 23, 24). Note that the flat energy band representation when excited materials have electrical charges jumps is a physically incorrect situation and model [7, 21, 39].

Based on the semiconductor type II band gap, there is no physical reason for the non absorption of most of the whole sunlight spectra leave's materials (chemical and physical structure), if photons from sunlight have energy equal or greater than both the A or B band gap's material. In this case, green leaves should not reflect "green" photons but absorb most of the sunlight spectra and emit green photons, through type II interface mechanism recombination, presented in figures 1, 2 and 3. Our optical excitation of green leaves shows green colour emission, when excited by 386 nm laser light (see SI-3). The same results are presented by Broglia and Chappelle[23, 24]. This is contrary to the postulated and presently accepted photosynthesis model. Most of the proposed photosynthesis mechanism models are based on energy transfer[1-5]. On SI-5 we compare some aspects of the presently accepted model and the model we propose for the photosynthesis first step mechanism. For the present type II interface model, utilizing staggered band gap, (e-, h+), carriers can move from staked leave's constituents as chromophores and so from energy levels. Part of these carries movement can releases energy but it is not based on energy transfer by resonance as it is proposed within the literature[1-5]. In the present model, we consider leaves' plants constituents (within a chloroplast) as a stack of multilayer below 10 nm thick as is the case for semiconductor artificial heterostructures[7-13, 21]. The proposed type II semiconductor interface model for the primary step of photosynthesis takes care of these colours/energy changes in nature's plants. In this case, nature's leaves with A and B band gap should change their physical and chemical properties with aging. So it does also for the A/B interface band gap alignment or overlap (ΔS), changing the colours to lower band gap energy emission. In other words, leave's plants aging deals with its chemical and physical parameters, changing the A/B allowed band gap and their ΔS relative position.

At the introduction we proposed some questions: to the question 'what colour was the Earth canopy for most of the present fossil petroleum we are using today', we could say that for the big trees and colossal forest, the spent energy to separate charges to get the photosynthesis process, the red shift is mostly the right answer. A picture proposing this red shift is presented on SI-6, for a 29% energy economy, for a sugar cane plantation, to workout its machinery and improve the energy production (biomass and sugar juice). The 29% economy is based on the $h\nu_{green}/h\nu_{red}$ quantum photon energy relationship. It should be mentioned here that few percent of this 29% economy applied to any of the present principal ethanol producers represents a big amount of a country economy for renewable energy applications. The renewable energy production efficiency increase could help developed countries to decrease $CO_2$ production. On Earth and extra-solar planets canopy studies, it could be better improved by considering the colour of leaves as being related to the spent energy for charge separation (or light emission) instead of simply a reflection without any relationship with its physical/chemical properties, as is the case today[1].

**Conclusions**

We present a couple of semiconductor material having a staggered band gap energy alignment, which has a very efficient interface light emission mechanism. Based on this emission mechanism, we propose that natural green leaves (and photosynthesis first step) band gap constituents can be associated to type II energy band gap. Most of the recent publications on optical research on chemicals from leaves from plants claim a quantum mechanics mechanism, without presenting qualitative or quantitative scientifically based arguments/mechanism/model. Ground state energy model and flat energy bands configuration are unrealistic energy band representations and it does not allow the development of a quantum mechanics mechanism. Furthermore, it does not allow the appearance of the necessary driving force for both (e-, h+) charges separation. The conclusion that green color is a reflection for plants is a non scientific statement taken from experiments that does not consider the natural chemical and physical environment of the leave's constituents. There is no experimental proof for the green color from intact leaves as being mostly a reflection. The above mentioned arguments answer why, at the present, we cannot explain the photosynthesis first step mechanism.

The type II interface recombination mechanism is a quantum mechanical effect by e- and h+ tunnelling/recombination/emission mechanism, based on their wavefunction overlap (barriers penetration). Here we explain why (e-, h+) attracting charges can be separated by using energy staggered configuration, under illumination.

This model represents an easier way to understand the photosynthesis primary mechanism through a staggered band gap line-up with a green colour energy being mostly an emission by the leave's plant at the interface of plant's chromophores constituents. The charges separation is a consequence of the interface recombination/emission by energy band bending. The energy band bending creating the necessary electric field to separate attracting electrical charges. The interface band bending originating from energetic offset's charges decaying, the leave's green colour emission can be considered as related to the spent energy to get charges' separation.

The authors are aware that the present proposal must be tested by experimental work. Maybe by having a proposed and existing mechanism it is easier to look for guidable experiments and reaching the goals, as compared to no mechanism at all or based on models that do not have any physical meaning.

Nature has created and improved the staggered energy configuration for materials that allows charges separation about 2-4 billions of years ago. Charges separation mechanism by illumination is at the origin of life on Earth. Human being appears about 200 millions of years ago; he lives within this mechanism using the two ends of this first step photosynthesis process: the biomass as food and oxygen to breath. Mother Nature has created an intriguing and interesting puzzle for the photosynthesis first step mechanism. It is time to discuss and understand the charges separation mechanism of the photosynthesis process. This mechanism needs acceptable physical conditions. If recent optical experiments results suggest a quantum mechanism for non-natural materials/medium studies, it does not suggest its association with the photosynthesis natural mechanism. The present proposals represent a new insight within the natural photosynthesis first step process.

**Acknowledgements**

The authors are grateful to the financial support from ANR-Filemon35-France, and the agencies Fundação de Amparo à Ciência e Tecnologia do Estado de Pernambuco (FACEPE) - Brazil and Conselho Nacional de Desenvolvimento Científico e Tecnológico (CNPq) - Brazil. The authors thank professors Cid Araújo, Celso P. Melo and Michael Sundheimer from DF-UFPE & UFRPE-Brazil for valuables discussions and suggestions.

***Support Information Available: SI-1 to 6***

## *Can the photosynthesis first step quantum mechanism be explained?*

*M. Sacilotti et al*

- Details of the calculations for the (e-, h+) wavefunction overlap at the AlInAs/InP interface : SI-1.
- Energy balance on type II interface : SI-2.
- Photoluminescence studies of green leaves : SI-3.
- The Mother's Nature energetic puzzle between few existing molecules extracted from leaves : SI-4, b.
- Comparing staggered energy band gaps and the presently accepted ground state energy model : SI-5
- Sugar cane plantation presented in the natural colour (green) and artificially red coloured, proposing $\approx 29\%$ of sunlight economy to workout its machinery and improve its efficiency for biomass production : SI-6.

# Support Information SI-1

Article proposed to Arxiv,

Title: **Can the photosynthesis first step quantum mechanism be explained?**

M. Sacilotti et al

Details of the calculations for the (e-, h+) wavefunction overlap at the AlInAs/InP interface.

In this work the bands are modelled as exponential bended bands, leading to Bessel wavefunctions for carriers.

Energy levels of carriers at valence and conduction bands, as well tunnelling length, for each kind of carrier are calculated by solving the transcendental equation obtained for wave function continuity at the interface.

*Parameters for the present calculation were taken from ref.* 39. Abraham, P. et al. Photoluminescence and band offsets of AlInAs/InP. *Semiconductor Sci. Technology* **10**, 1-10 (1995).

Both bands of valence and conduction bands have similar, bent on account of them injected carriers. In each of the potential will be approximated by exponential functions as shown in eq. (1) and picture in Fig. 1.

$$\begin{aligned} V(x) &= V_o + \delta\left(1 - e^{-x/\sigma}\right) \quad x \geq 0 \\ V(x) &= V_1 - \Delta\left(1 - e^{x/\eta}\right) \quad x < 0 \end{aligned} \quad (1)$$

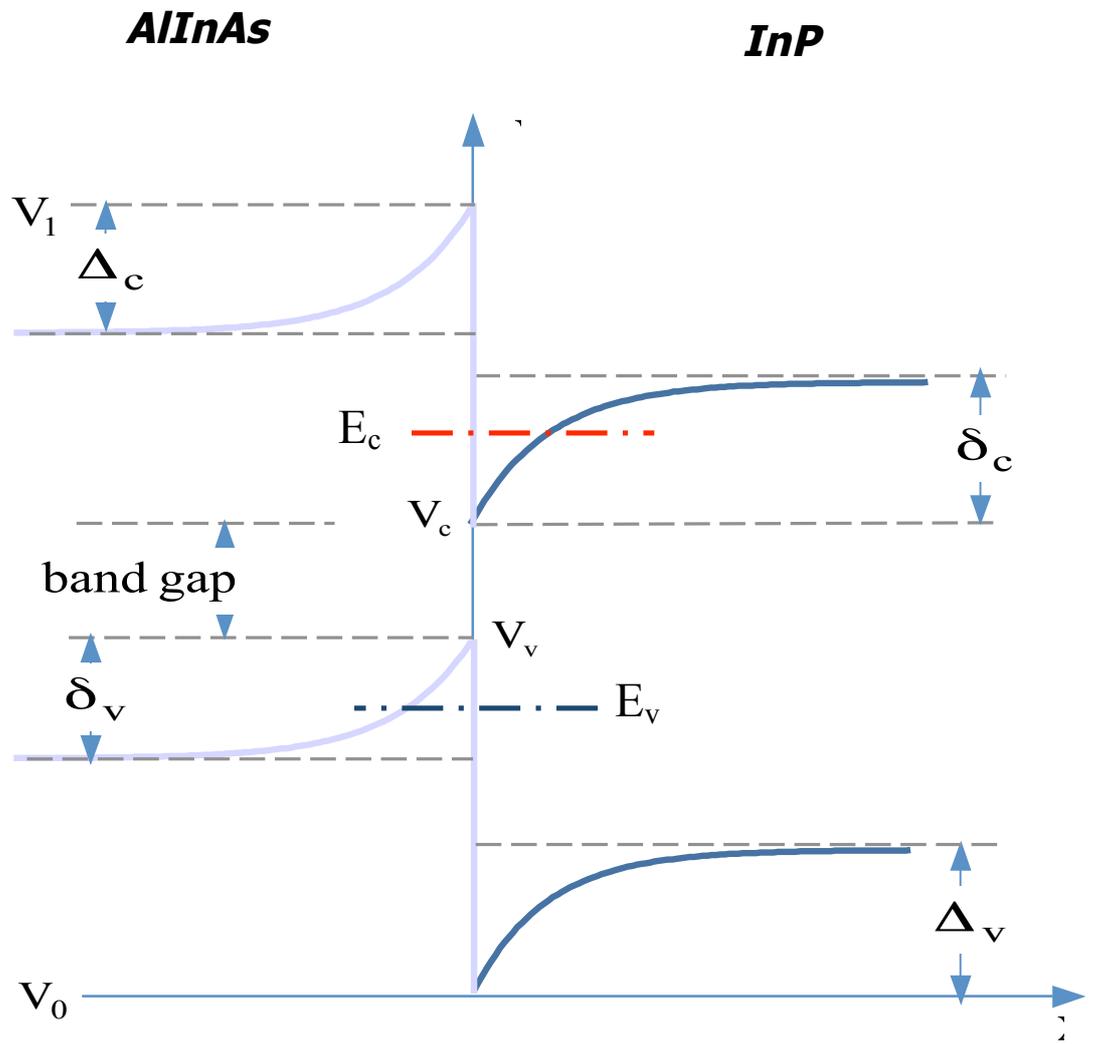

Fig. 1 – The scheme of tilted electronic bands for AlInAs/InP.

Substituting these potentials in the Schroedinger equation we have:

$$\frac{d^2\psi}{dx^2} + \frac{2m^*}{\hbar^2}\left\{E - \left[V_o + \delta\left(1 - e^{-x/\sigma}\right)\right]\right\}\psi = 0 \quad x \geq 0$$
$$\frac{d^2\psi}{dx^2} + \frac{2m^*}{\hbar^2}\left\{E - \left[V_1 - \Delta\left(1 - e^{x/\eta}\right)\right]\right\}\psi = 0 \quad x < 0 \quad (2)$$

These equations are applied to both the calculation of energy levels for the holes in the valence band and for the electrons in the conduction band. In each case we should observe the proper values of the parameters of the table 1. Using the transformation of variables

$$\xi_+ = \left(2\sigma\sqrt{\frac{2m^*}{\hbar^2}\delta}\right)e^{-x/2\sigma}$$
$$\xi_- = \left(2\eta\sqrt{\frac{2m^*}{\hbar^2}\Delta}\right)e^{x/2\eta} \quad (3)$$

We get the following equations:

$$\xi_+^2 \frac{d^2\psi}{d\xi_+^2} + \xi_+ \frac{d\psi}{d\xi_+} + \left(\xi_+^2 - 4\nu^2\right)\psi = 0 \quad x \geq 0$$
$$\xi_-^2 \frac{d^2\psi}{d\xi_-^2} + \xi_- \frac{d\psi}{d\xi_-} - \left(\xi_-^2 + 4\mu^2\right)\psi = 0 \quad x < 0 \quad (4)$$

These are the known Bessel equations in which it is defined the parameters:

$$\nu^2 = \frac{2m^*\sigma^2}{\hbar^2}\left[(V_o + \delta) - E\right]$$
$$\mu^2 = \frac{2m^*\eta^2}{\hbar^2}\left[(V_1 - \delta) - E\right] \quad (5)$$

The solutions for the eqs. (4) will be:

$$\Psi_+(z) = AJ_{2\nu}(\xi_+) \quad x \geq 0$$
$$\Psi_-(z) = BI_{2\mu}(\xi_-) \quad x < 0 \quad (6)$$

These solutions must satisfy the boundary conditions at z = 0. Equating the functions and their derivatives at z = 0 results the transcendental equation:

$$\frac{\nu}{\sigma}\frac{J_{2\nu+1}(2\nu)}{J_{2\nu}(2\nu)} = \left(\frac{\sigma\mu}{\nu\eta} + \frac{\nu}{\sigma}\right) + \frac{\sigma u}{\nu\eta}\frac{I_{2\mu+1}(2u)}{I_{2\mu}(2u)} \quad (7)$$

The parameters concerning the potential wells in the conduction band and valence are those given in Table 1.

| $V_o$ (meV) | $V_1$ (meV) | $\delta$ (meV) | $\Delta$ (meV) | $\sigma$ (nm) | $\eta$ (nm) | $m_e$ |
|---|---|---|---|---|---|---|
| 0 | 349 | 60 | 20 | 5 | 5 | 0,07 |
| 0 | 272 | 60 | 20 | 5 | 5 | 0,4 |
| Table 1 | | | | | | |

Solving the transcendental equations for the valence and conduction bands it was obtained two energy levels for holes (VB) and one energy level for electrons (BC) in the conduction band. The energies and the parameters that determine the wave functions for electrons and holes are given in Table 2.

| | A | B | E (meV) | $V_1$-E (meV) |
|---|---|---|---|---|
| Electrons | 0,491 | 804,794 | 54,68 | 5,32 |
| Holes | 1,223 | $5,313 \times 10^5$ | 38,85 | 232,68 |
| | 0,689 | $-9,9 \times 10^4$ | 56,13 | 215,87 |
| Table 2 | | | | |

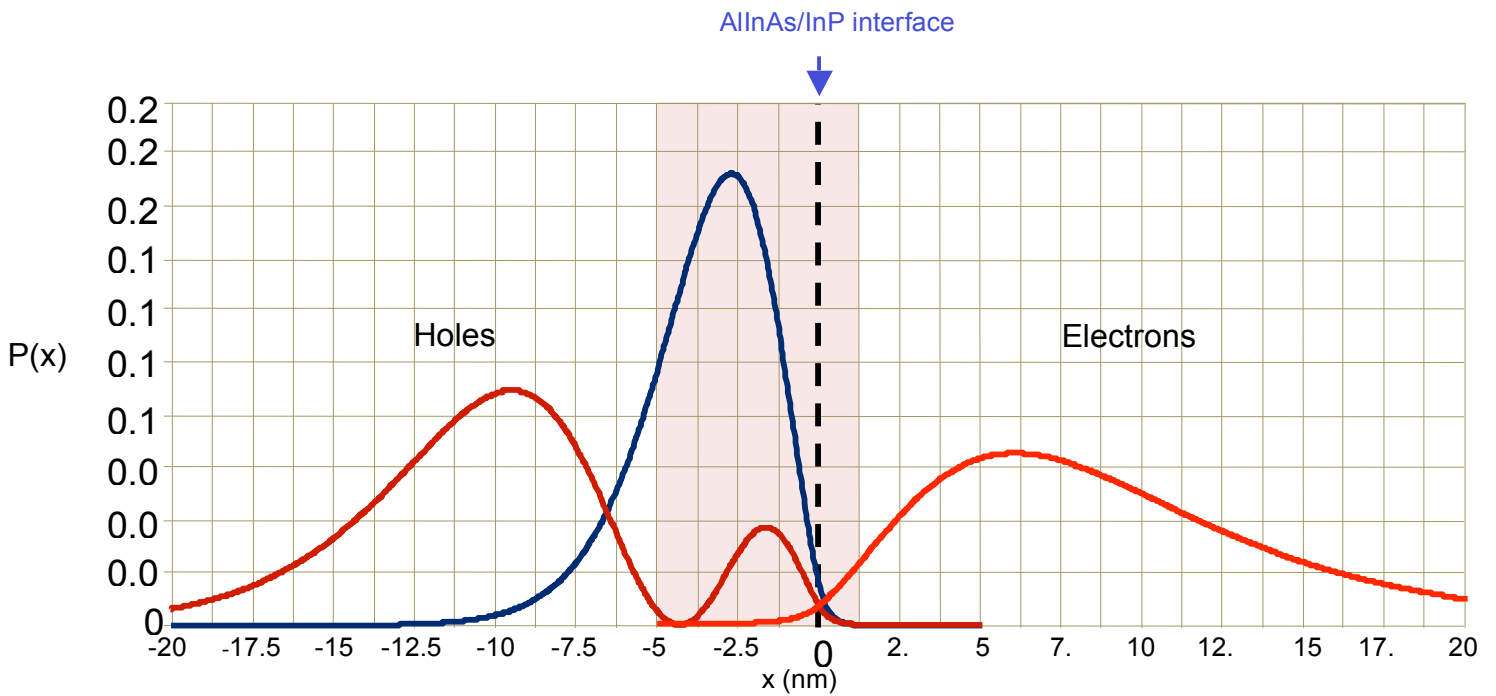

Fig. 2 It shows the probability density versus distance within the structure, for the valence band and conduction band, where we see that the penetration probability for electrons and holes beyond the wall of the well. The region with interpenetrating wavefunctions is pointed out in the figure by pink colored rectangle.

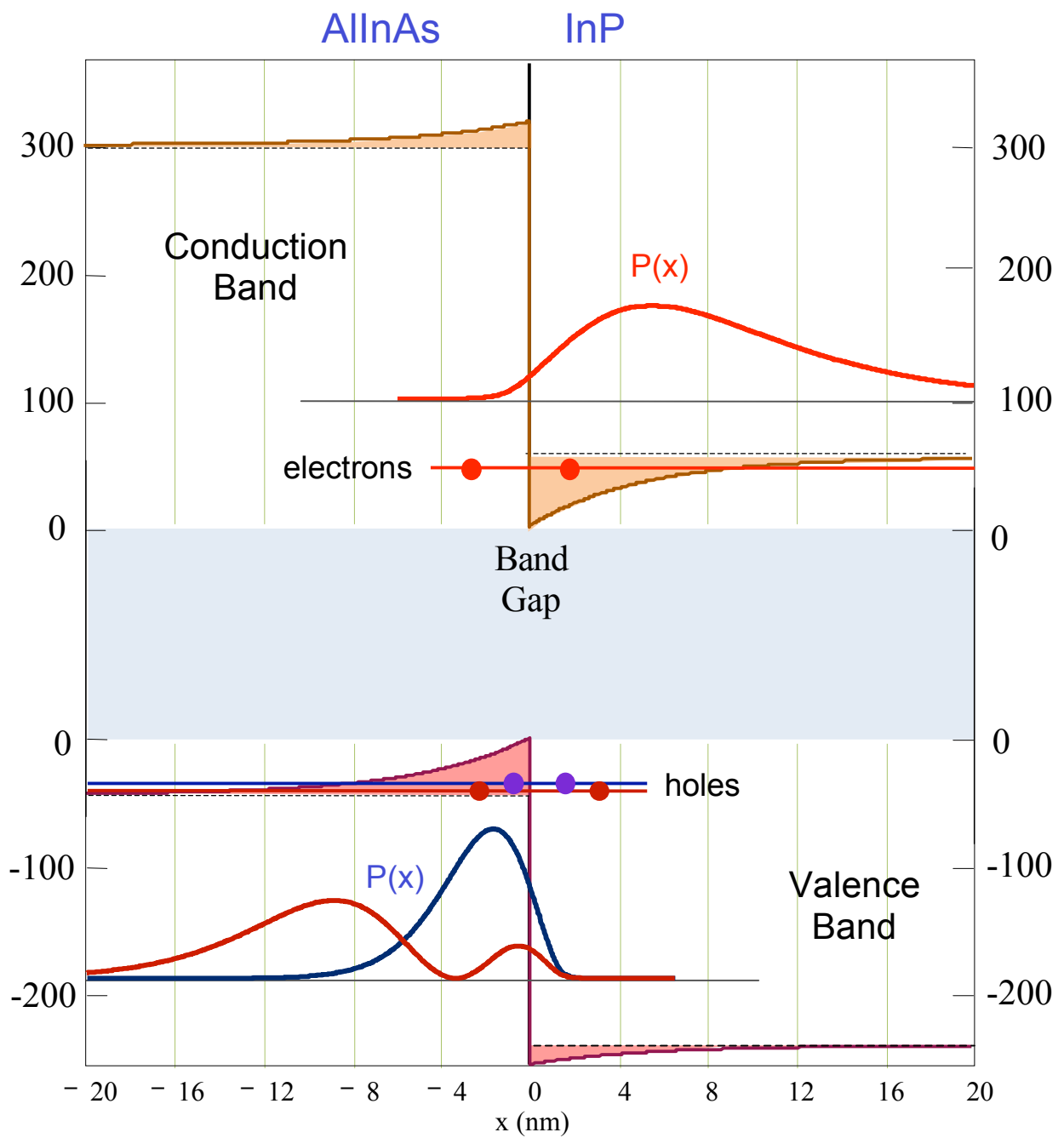

Fig. 3 It shows schematically, with the values of the energy-scale the probability densities for electrons and holes relative to the potential wells within the system AlInAs/InP. x (nm) is the distance from the interface.

SI-2

## Energy balance on type II interface

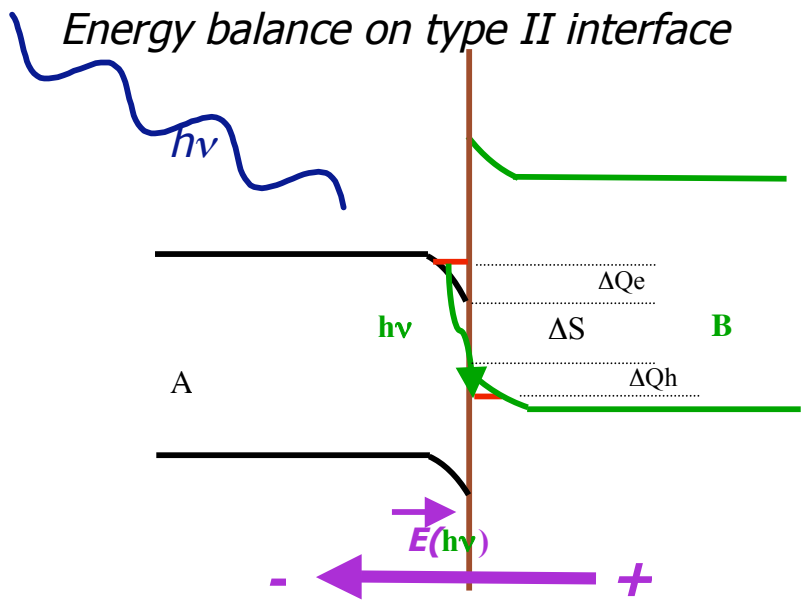
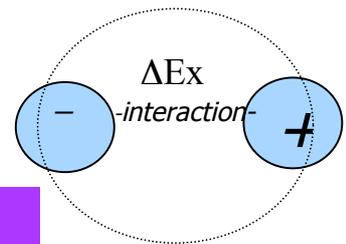

**The driving force for the (e⁻, h⁺) separation**

Interface light emission energy : $h\nu = \Delta S + \Delta Qe + \Delta Qh - \Delta Ex$

Material B band gap energy > $h\nu$ < Material A band gap energy

Where $\Delta S$ = A/B band energy overlap, $\Delta Qe$ = electron energy quantisation, $\Delta Qh$ = hole energy quantisation and $\Delta Ex$ = electron-hole energy exciton.

**SI-3**

**Photoluminescence studies of green leaves.**

We performed optical absorption and photoluminescence (PL) experiments of leaves at room temperature to see if there are emissions from them. We used intact young yellow-green *Erytrina indica picta* leaves exposed to green (532 nm) and violet (386 nm) laser light excitation. This leaf is naturally composed of green and yellow regions (figures SI-3-1 to 3). The green region of the leaves exposed to these excitation wavelengths resulted in broadband light emission on the blue, green, yellow and red spectral regions.

Figure SI-3-2 presents the PL experimental results for the 532 nm excitation, by using a 512-560 nm filter for detection and avoid the excitation peak. The green part of the leaf presents a broad peak at the red side of the spectrum (675-825 nm). The yellow part of the leaf presents two main peaks (orange colour) between 560-750 nm. So, the 532 nm excitation gives red shift light emission. Leaf excitation by short wavelength light gave green light as emission (see SI-3-3). As there is no green light excitation for the 386 nm we can come to a controversial conclusion that the green color that our eyes see is mostly an emission and it is contrary to what we learn in schools and specialized literature[1-5]. Note that these experiments were performed *in vivo* and not on chemicals separated from leaves. The conversion mechanism from ultra-violet, violet and blue excitation to green, yellow and red emission is postulated as primarily due to type II energetic interfaces. This statement does not imply that there is no green (and others colours) reflection, transmission, scattering and diffusion when leaves are excited with sunlight. Also, it cannot be ruled out the possibility of green absorption and red shift green emission by the many possibilities of existing type II interface within a leaf (see SI-4).

Blue-green laser induced fluorescence from intact leaves has been observed by many research groups, using UV (308 nm) excitation as M. Broglia and E. Chapelle et al [23, 24]. There is no conclusion presented by these authors, for the observed green fluorescence. Their results show also that the blue-green-red fluorescence can be ascribed to others structures, such as cell envelopes and vacuolar solutes [23]. This indicates that the environment of the constituents of the leaves play an important role in colour emission from leaves. Nanosecond decay of chlorophyll fluorescence from leaves has been attributed to the recombination of separated charges in the reaction centre of PSII[25-27]. The environment of the leaves' constituents and the recombination of separated charges, both support our proposition for green light emission as being a type II interface mechanism, as separated charges.

As most of the leaves contain many kinds of molecules having 3-5 absorption peaks, when separated from the matrix, it is reasonable to expect many possible ΔS energy band overlaps from these absorption bands when they are not separated from the leaf (or in intact leaves).

On SI-4 we present one of these type II interfaces possibilities for many chemicals green leaves constituents.

## SI-3-1

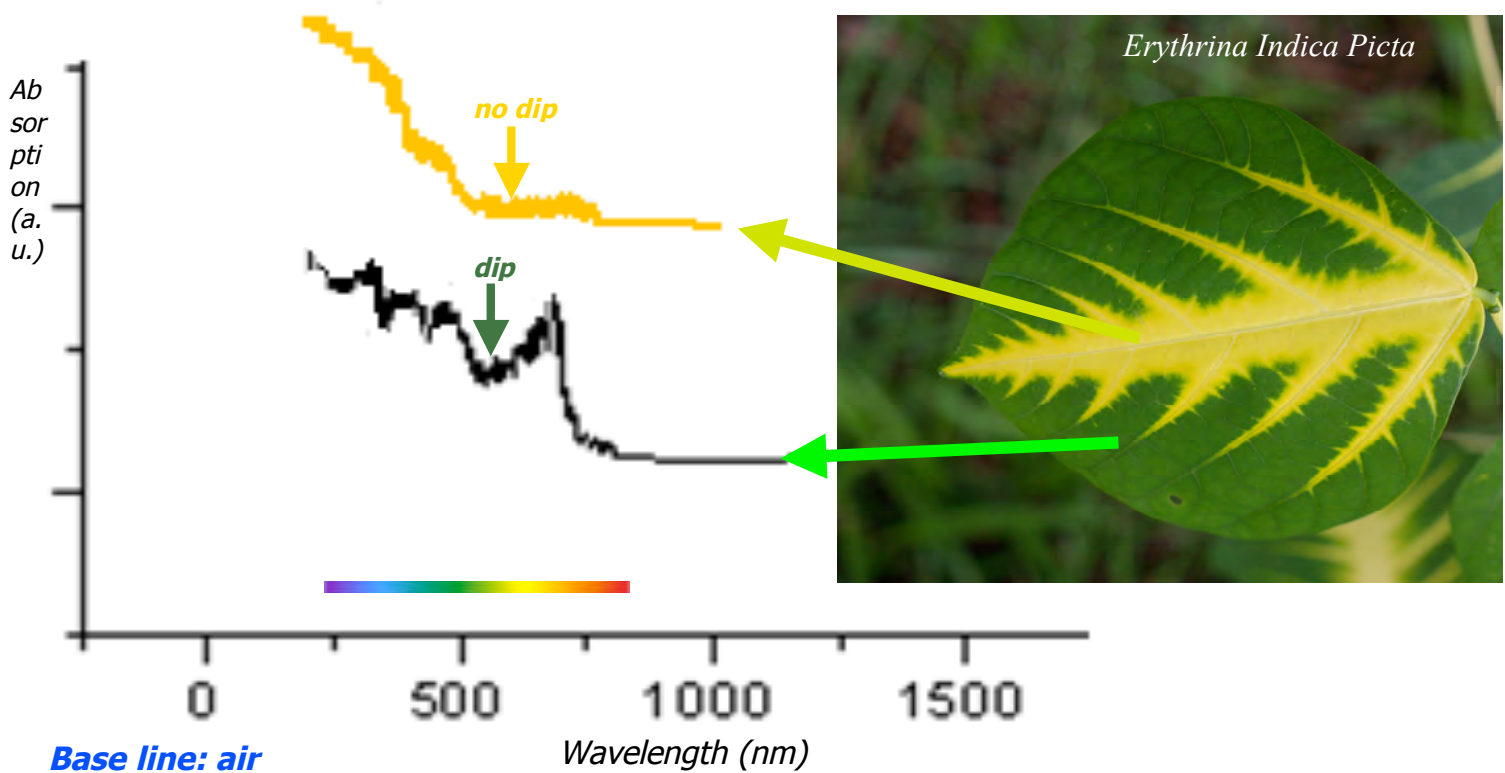

*Figure SI-3-1 This figure represents the absorption x wavelength for a leaf. Room temperature. This is the usual behaviour for natural leaves, with a deep within the green spectral region. Equipement: Cary 50E*

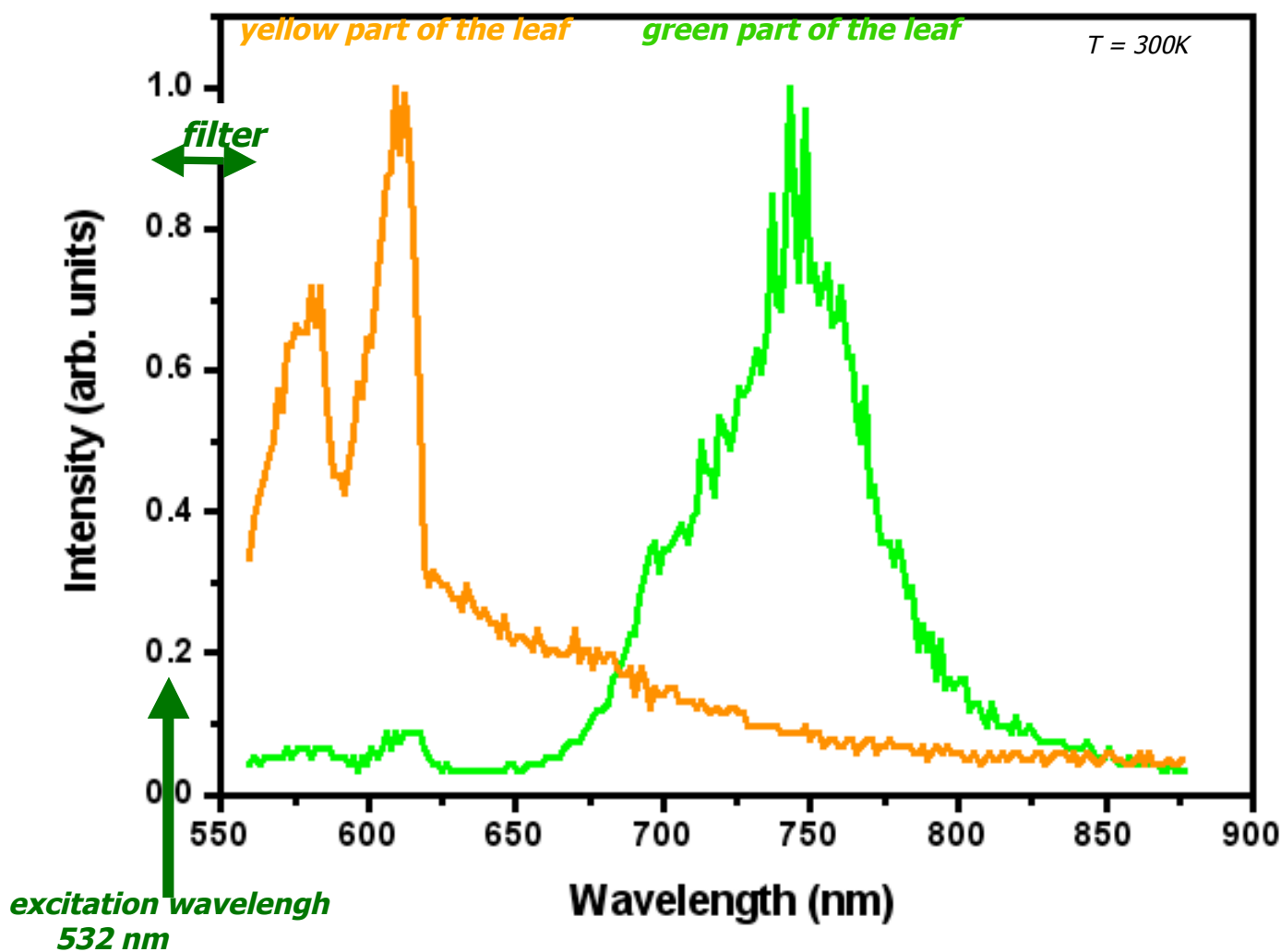

**Figure SI-3-2**. Room temperature photoluminescence spectra of yellow-green intact *Erytrina indica picta* leaves, excited with 532nm wavelengh, 10 ns pulses from a Nd:YAG laser. To avoid the 532 nm laser line, a 532-560 nm filter is placed between the leaf and the spectrometer. Orange spectrum is from the yellow part of the leaf. Green colour spectrum is from the green part of the leaf. No correction is proposed for the intensity axe. No smoothing was performed for both curves. These two spectra suggest that the emission spectra of different regions (colours) in a leave depends also on the excitation wavelengths.

*SI-3-3*

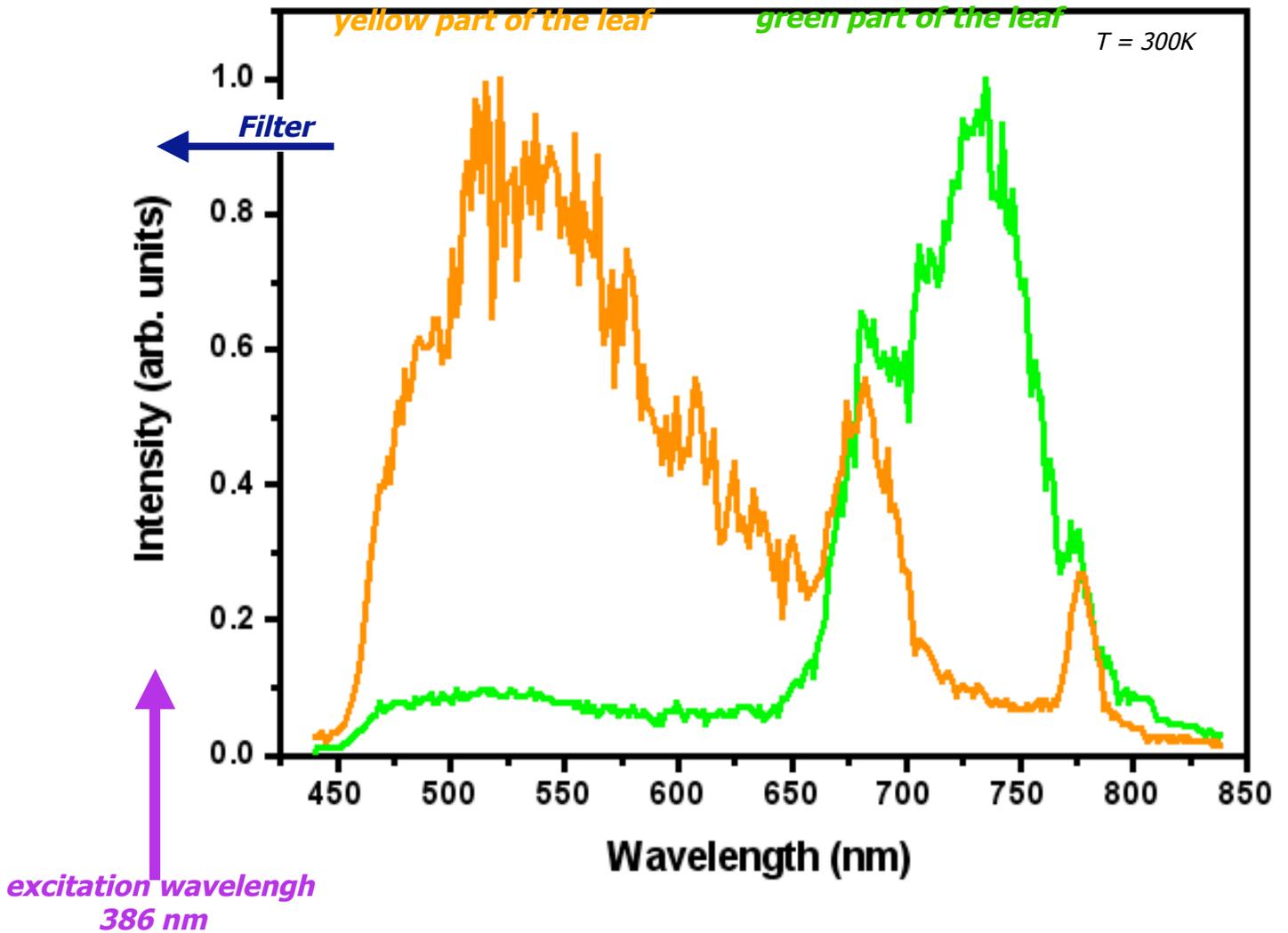

Figure SI-3-3. Room temperature photoluminescence spectra of yellow-green intact *Erytrina indica picta* leaves, excited with 386 nm wavelengh, 10 ns pulses from a Nd:YAG laser. To avoid the 386 nm laser line, a cutoff filter below 450 nm is placed between the leaf and the spectrometer. Orange spectrum is from the yellow part of the leaf. Green colour spectrum is from the green part of the leaf. No corrections is proposed for the intensity axe. No smoothing was performed for both curves. These two spectra suggest that the emission spectra of different regions (colours) in a leave depends also on the excitation wavelengths.

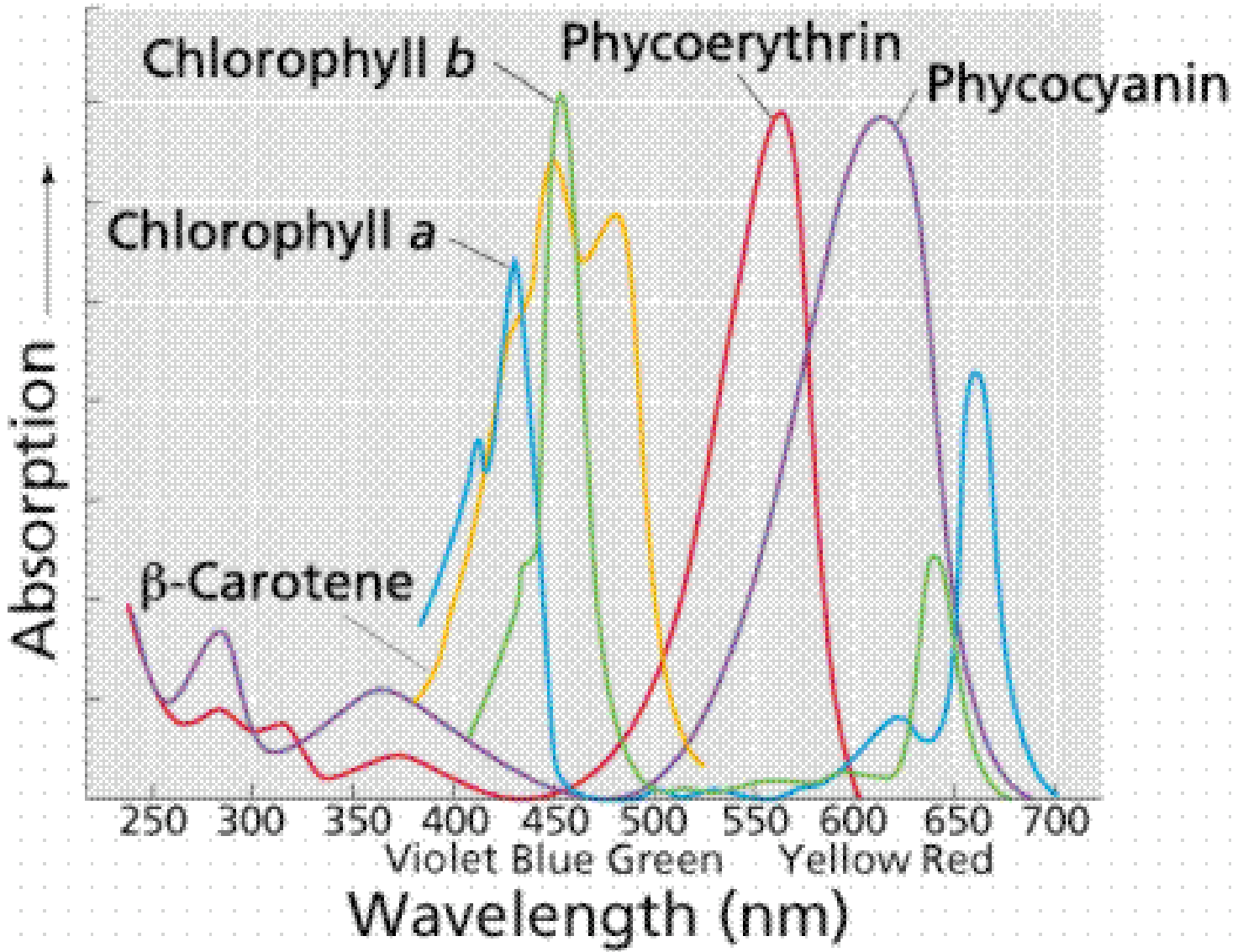

**Figure SI-4a.** Picture representing the energy absorption peaks for chemicals from plants (taken from the literature). In a real situation (*in vivo*), these absorption peaks should change their positions and intensities. It is not excluded the possibility to find out new absorption peaks for the natural *in vivo* plants. These absorptions peaks were utilised to build the figure SI-4b below.

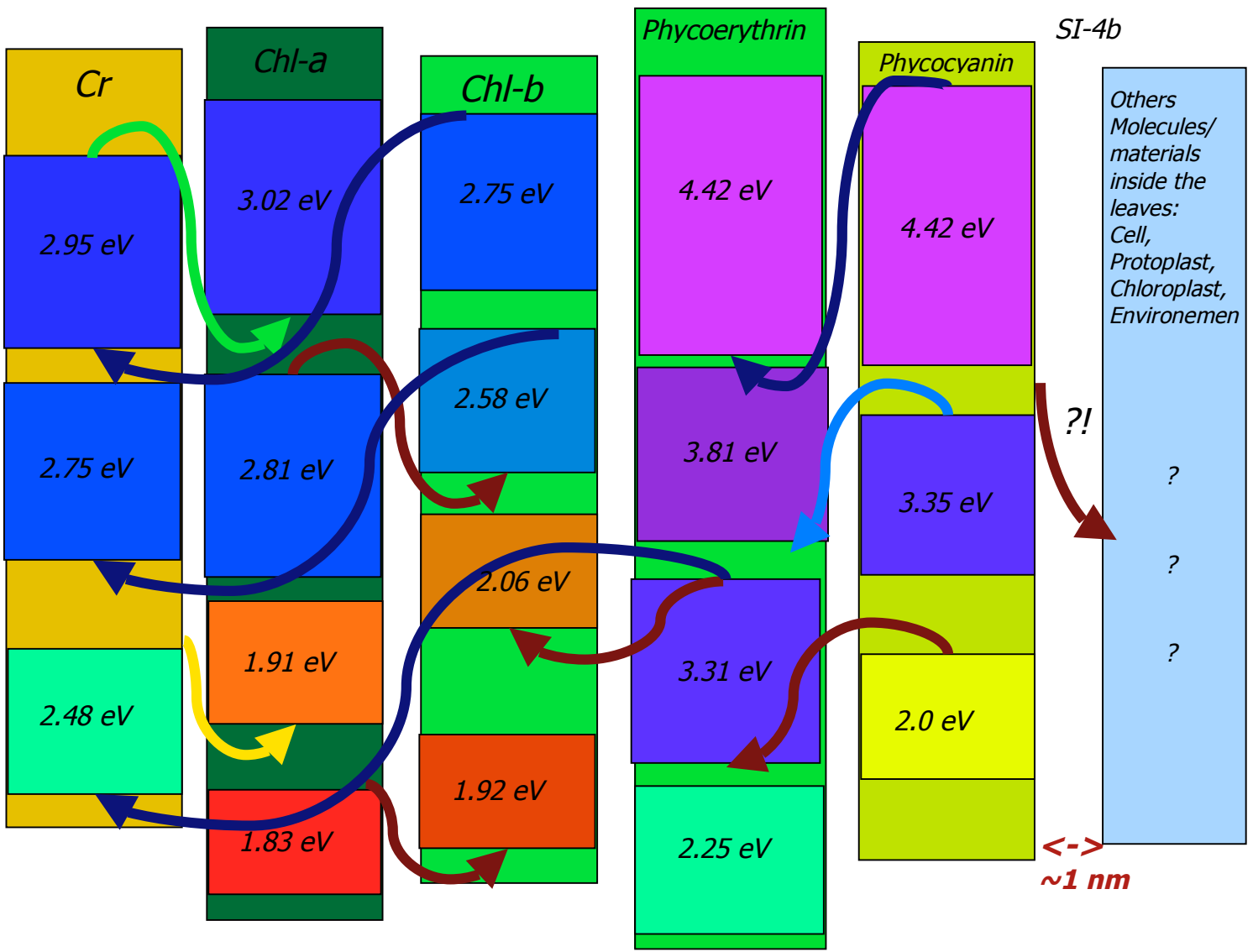

**Figure SI-4.** The Mother Nature puzzle.

Representation (on strips) of energy band gap for the leaves' constituent, containing 3-5 mains peaks absorption depicted within each strip. These energy bands absorption were taken from the literature. The energy bands absorption for the cell, protoplast and chloroplast environment are not know by the authors. From one of these band gaps energy relative position (a staggered one) emerge the main green colour of leaves we have today. From others energy relative's positions emerge the others less intense colours as yellow, brown, red…, from leaves. Genetically leaves can use them on machinery for adaptation/evolution and have different colours emission. **eV** represents the electron volts energy from the absorption energy peaks, extracted from the literature. Note also that these energy absorption values were taken from chemical separated from the leaves. For in vivo leaves it should change these values because they are interface related. Note that these strips configuration is the only one in between many others possibilities to have type II (staggered) interface. This figure SI-4 shows that, even for a non in vivo situation, it is impossible not to have a staggered band gap energy configuration between two or more molecules. **Cr** = carotene, **Chl** = chlorophyll. **~ 1 nm** represents the size of and distances in between molecules within a leaf. **?** represents unknown band gap energy for others materials within a leaf. **?!** represents unknown band gap energy relative position between these materials and Cr, Chl, Phycoerythrin, Phycocyanin, etc. The staggered energy configuration gives us the physical situation to have a quantum mechanism for photosynthesis. The arrows indicate possible staggered energy band gap relative position and (e-, h+) recombination/transitions. As (e-, h+) are seated on different materials, quantum k selections rules do not hold anymore for these recombination/transitions.

## SI-5 Comparing staggered energy band gaps and the presently accepted ground state energy model

The currently accepted model for the photosynthesis mechanism is based roughly on absorption of visible sunlight in the red and blue regions of the spectra, with no absorption (and consequently a conclusion to the existence of a reflection and/or scattering mechanism) in the green spectral region[1-5, 28-31]. The present proposed model for photosynthesis, based on type II energetic interfaces, is based on the sun visible light spectra absorption by the Nature's green leaves constituent. The green colours of the Nature's leaves come from light emission by the (e-, h+) recombination from both materials, type II interface, or e- seated in material A and the h+ seated in material B as represented in figure SI-2. These materials A, B, C, D, … can be chlorophylls, carotenoids, phycoerythrin, phycocyanin, protoplast, water, etc and the physical structure within a chloroplast. Most of the Nature's leaves get colours like yellow/brown/red (decreasing energy of the observed spectra) when they get older (Spring for cold countries). To this point, a remark can be made on aging of leaves: red shift (lower energy) in green plant's leaves is an intelligent way of energy economy. Otherwise, a blue shift of its colours (higher energy) is a no sense tendency.

The green (or yellow or red) colours of Nature's leaves depends on the (e-, h+)

wavefunction overlap or tunnelling process. It is a very fast process (compared with usual CB-VB semiconductor recombination mechanism) and its efficiency comes from the non existing quantum mechanics selection rules, even if it is a quantum mechanics assisted mechanism as above proposed.

Recently many papers based on very sophisticated and beautiful optical experiments got to the conclusion that photosynthesis is a quantum mechanical mechanism, but without presenting it[3, 4, 22, 28, 32-35]. The present staggered band gap energy proposal supports this quantum mechanical dynamic mechanism. F. J. Schmitt et al[36], by using optical measurements of chlorophyll get to the conclusion of a double well mechanism. Our proposal for a staggered energetic representation fits exactly into this double well conclusion. It goes further: both well are separated by an interface between two different materials. Based on theoretical calculations, De Angelis et al[37] got to the conclusion for a staggered energy representation between two organic molecules, the same concept holds for inorganic/organic materials[37, 38]. That is exactly what we are proposing: (e-, h+) charges have energetic steps when travelling from a material to another. This physical energetic steps concept cannot be put beside and it is the basis for when working and considering different materials. In another way, electric charges have always energetic steps when moving from a material to another. On the contrary, the ground state energy representation (or a band gap engineering, old of tens of years) is an unrealistic and non existing physical interface energetic representation[1-5, 7-13]. The ground state energy does not allow a quantum mechanics mechanism development for the photosynthesis first step explanation[3, 4, 7, 28, 32-35, 37]. There is no proof of and for the "ground stage energy" representation. It does not allow the apparition of an electric field for both (e-, h+) charges separation.

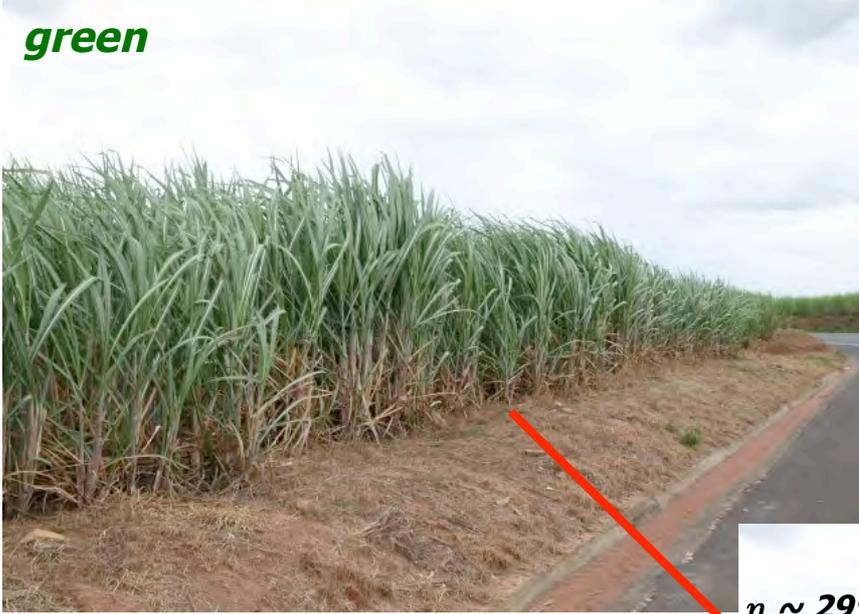

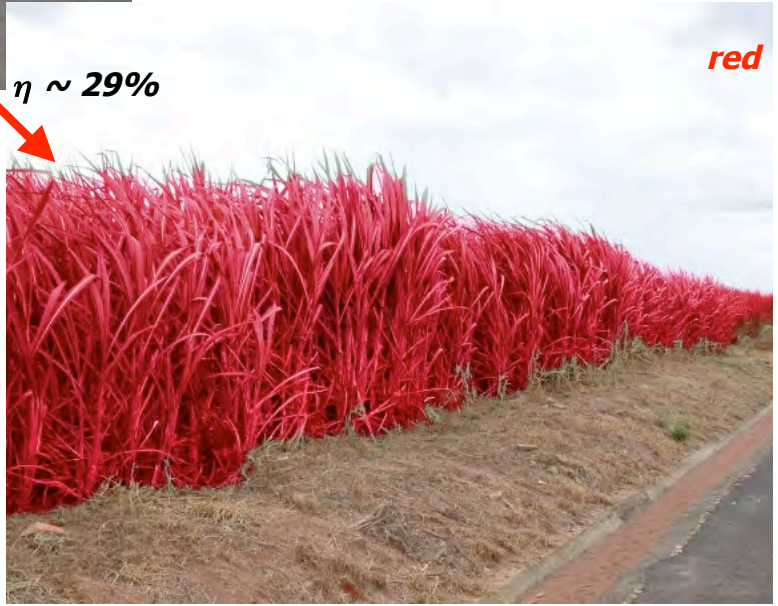

*Figure SI-6. Sugar cane plantation presented in the natural colour (green) and artifitially red coloured.* ($h\nu_{green}$ - $h\nu_{red}$)/$h\nu_{green}$ ≈ 29% represents energy economy for the sugar cane machinery to work out its on processes and produce more biomass and/or sugar, if the colour can be related to an emission for the photosynthesis mechanism and related to the spent energy to separate attracting electrical charges.